\documentclass[prd,amsmath,amssymb,showpacs,superscriptaddress,nofootinbib,twocolumn]{revtex4-1}

\usepackage{graphicx}
\usepackage{epsfig,graphics,subfigure,psfrag,amsmath,amssymb}
\usepackage{lineno}
\usepackage{dcolumn}
\usepackage{bm}
\usepackage{overpic}

\usepackage[english]{babel}
\addto{\captionsenglish}{%

}

\newcommand{\BR}{{\cal B}}

\newcommand{\jpsi}{J/\psi}

\usepackage[
total={6.5in,8.75in}, top=1.2in, left=0.9in
]{geometry}

\begin{document}

\title{
\boldmath  Observation of the electromagnetic doubly OZI-suppressed
decay $\jpsi \rightarrow \phi\pi^{0}$
}

\newcommand{\piz}{\pi^{0}}
\newcommand{\Kp}{K^{+}}
\newcommand{\Km}{K^{-}}
\newcommand{\NobsA}{838.5 \pm 45.8}
\newcommand{\NobsB}{35.3 \pm 9.3}
\newcommand{\phaseA}{-95.9^\circ \pm 1.5^\circ}
\newcommand{\phaseB}{-152.1^\circ \pm 7.7^\circ}
\newcommand{\solutionAs}{(2.94 \pm 0.16) \times 10^{-6}}
\newcommand{\solutionBs}{(1.24 \pm 0.33) \times 10^{-7}}
\newcommand{\solutionA}{[2.94 \pm 0.16\text{(stat.)} \pm 0.16\text{(syst.)}] \times 10^{-6}}
\newcommand{\solutionB}{[1.24 \pm 0.33\text{(stat.)} \pm 0.30\text{(syst.)}] \times 10^{-7}}

\author{
  \begin{small}
    \begin{center}
      M.~Ablikim$^{1}$, M.~N.~Achasov$^{9,a}$, X.~C.~Ai$^{1}$,
      O.~Albayrak$^{5}$, M.~Albrecht$^{4}$, D.~J.~Ambrose$^{44}$,
      A.~Amoroso$^{48A,48C}$, F.~F.~An$^{1}$, Q.~An$^{45}$,
      J.~Z.~Bai$^{1}$
% ...
      \\
      \vspace{0.2cm}
      (BESIII Collaboration)\\
      \vspace{0.2cm} {\it
        $^{1}$ Institute of High Energy Physics, Beijing 100049, People's Republic of China\\
% ...
$^{4}$ Bochum Ruhr-University, D-44780 Bochum, Germany\\
        $^{5}$ Carnegie Mellon University, Pittsburgh, Pennsylvania 15213, USA\\
% ...
$^{9}$ G.I. Budker Institute of Nuclear Physics SB RAS (BINP), Novosibirsk 630090, Russia\\
% ...
$^{44}$ University of Rochester, Rochester, New York 14627, USA\\
        $^{45}$ University of Science and Technology of China, Hefei 230026, People's Republic of China\\
% ...
$^{48}$ (A)University of Turin, I-10125, Turin, Italy; (B)University of Eastern Piedmont, I-15121, Alessandria, Italy; (C)INFN, I-10125, Turin, Italy\\
% ...
$^{a}$ Also at the Novosibirsk State University, Novosibirsk, 630090, Russia\\
% ...
      }\end{center}
    \vspace{0.4cm}
  \end{small}
}
\affiliation{}

\begin{abstract}
Using a sample of $1.31$ billion $\jpsi$ events accumulated with the BESIII
detector at the BEPCII collider, we report the observation of the
decay $\jpsi \rightarrow \phi\piz$, which is the first evidence for a
doubly Okubo-Zweig-Iizuka suppressed electromagnetic $\jpsi$ decay. A
clear structure is observed in the $\Kp \Km$ mass spectrum around 1.02~GeV/$c^2$, which can be
attributed to interference between $\jpsi \rightarrow \phi\piz$ and
$\jpsi \rightarrow \Kp\Km\piz$ decays. Due to this interference, two possible solutions are found. The corresponding
measured values of the branching fraction of $\jpsi \to \phi\piz$ are $\solutionA$ and $\solutionB$.

\end{abstract}

\maketitle

The discovery of the $J/\psi$ played an important role in
understanding the basic constituents of nature and opened a new era in
particle physics. Its unexpected narrow decay width provided insight
into the study of strong interactions. As its mass is below the charmed
meson pair threshold, direct decay into charmed mesons is
forbidden. Therefore the $\jpsi$ hadronic decay modes are
Okubo-Zweig-Iizuka (OZI)~\cite{OZI} suppressed, and the final states
are composed only of light hadrons.

A full investigation of $\jpsi$ decaying to a vector meson ($V$) and a
pseudoscalar meson ($P$) can provide rich information about SU(3) flavor
symmetry and its breaking, probe the quark and gluon content of the
pseudoscalar mesons, and determine the electromagnetic
amplitudes~\cite{Haber,Seiden,Rafel}. However the presence of doubly
OZI (DOZI) suppressed processes, like the observation of $J/\psi$
radiatively decaying into $\omega\phi$~\cite{bes2wphi,bes3wphi}, complicates matters as
they do not obey quark correlation or satisfy nonet symmetry (treating SU(3) octets and singlet
as a nonet and assuming the coupling constants are the same in the interactions~\cite{Haber,Seiden}).
Well established phenomenological models~\cite{Haber,Seiden} have
indicated that the DOZI amplitude can have a large impact through
interference with the singly OZI suppressed amplitude.

Of interest is the decay $\jpsi \to \phi \piz$, which occurs via the
electromagnetic DOZI process or by non-ideal $\omega-\phi$
mixing~\cite{Haber,Seiden,MARKIII}. Recently, using a
combination of a factorization scheme for the strong decays and a
Vector Meson Dominance (VMD) model for electromagnetic decays in
$J/\psi\rightarrow VP$, the branching fraction  of
$J/\psi\rightarrow\phi\piz$ has been predicted to be around $8\times
10^{-7}$~\cite{Zhao}, while the best upper limit to date comes from
the BES collaboration,
$\mathcal{B}(J/\psi\rightarrow\phi\piz)<6.4\times 10^{-6}$ at the
90\% confidence level (C.L.)~\cite{bes2phipi0}. In this paper, we
report the first observation of $\jpsi \to \phi \piz$ based on a
sample of $(1.311 \pm 0.011) \times 10^{9}$ $\jpsi$
events~\cite{Njpsi09,Njpsi12} accumulated with the BESIII
detector.

The BESIII detector~\cite{BES} is a magnetic spectrometer located at
the Beijing Electron Positron Collider (BEPCII), which is a
double-ring $e^+e^-$ collider with a design peak luminosity of
$10^{33}$ cm$^{-2}$s$^{-1}$ at the center of mass (c.m.) energy of
$3.773$~GeV. The cylindrical core of the BESIII detector consists of a
helium-based main drift chamber (MDC), a plastic scintillator
time-of-flight system (TOF), and a CsI(Tl) electromagnetic calorimeter
(EMC). All of them are enclosed in a superconducting solenoidal magnet
providing a $1.0$ T ($0.9$ T in 2012) magnetic field. The solenoid is
supported by an octagonal flux-return yoke with resistive plate
counter muon identifier modules interleaved with steel. The acceptance
for charged particles and photons is $93\%$ of $4\pi$ solid angle. The
charged-particle momentum resolution is $0.5\%$ at $1$~GeV/$c$, and
the specific energy loss ($dE/dx$) resolution is $6\%$. The EMC measures photon energies with
a resolution of $2.5\%$ (5\%) at 1~GeV in the barrel (endcaps). The
time resolution of the TOF is $80$ ps in the barrel and $110$ ps in the
endcaps. The BESIII offline software system (BOSS) framework is based
on Gaudi~\cite{Gaudi}. A GEANT4-based \cite{Geant4} Monte Carlo (MC) simulation is
used to determine detection efficiencies and estimate backgrounds.

For the decay $\jpsi \to \phi \piz \to \Kp \Km \gamma\gamma$,
a candidate event is required to have two charged tracks with opposite
charge and at least two photons. For each charged track, the polar
angle in the MDC must satisfy $|\cos \theta|<0.93$, and the point of
closest approach to the $e^+e^-$ interaction point must be within
$\pm10$ cm in the beam direction and within $1$ cm in the plane
perpendicular to the beam direction. TOF and $dE/dx$ information are combined to give particle identification
(PID) probabilities for $\pi$, $K$ and $p$ hypotheses. To identify a track as a
kaon, the PID probability for the kaon hypothesis must be larger
than that for the pion hypothesis.

For each photon, the energy deposited in the EMC must be at least $25$
MeV for $|\cos \theta|<0.8$ or $50$~MeV for
$0.86<|\cos\theta|<0.92$. To select isolated showers, the angle
relative to the nearest charged track must be larger than
$20^\circ$. The timing information of the EMC is used to suppress
electronic noise and unrelated energy deposits. Furthermore, a
four-constraint (4C) kinematic fit is applied to the candidate events
under the $\Kp \Km \gamma\gamma$ hypothesis, requiring the 4-momentum
of the final state to be equal to that of the colliding beams. If
there are more than two photon candidates in an event, the combination
with the smallest $\chi_{4C}^2(\Kp\Km\gamma\gamma)$ is
retained. Events with $\chi_{4C}^2<30$ are selected.

\begin{figure}
\includegraphics[width=0.4\textwidth]{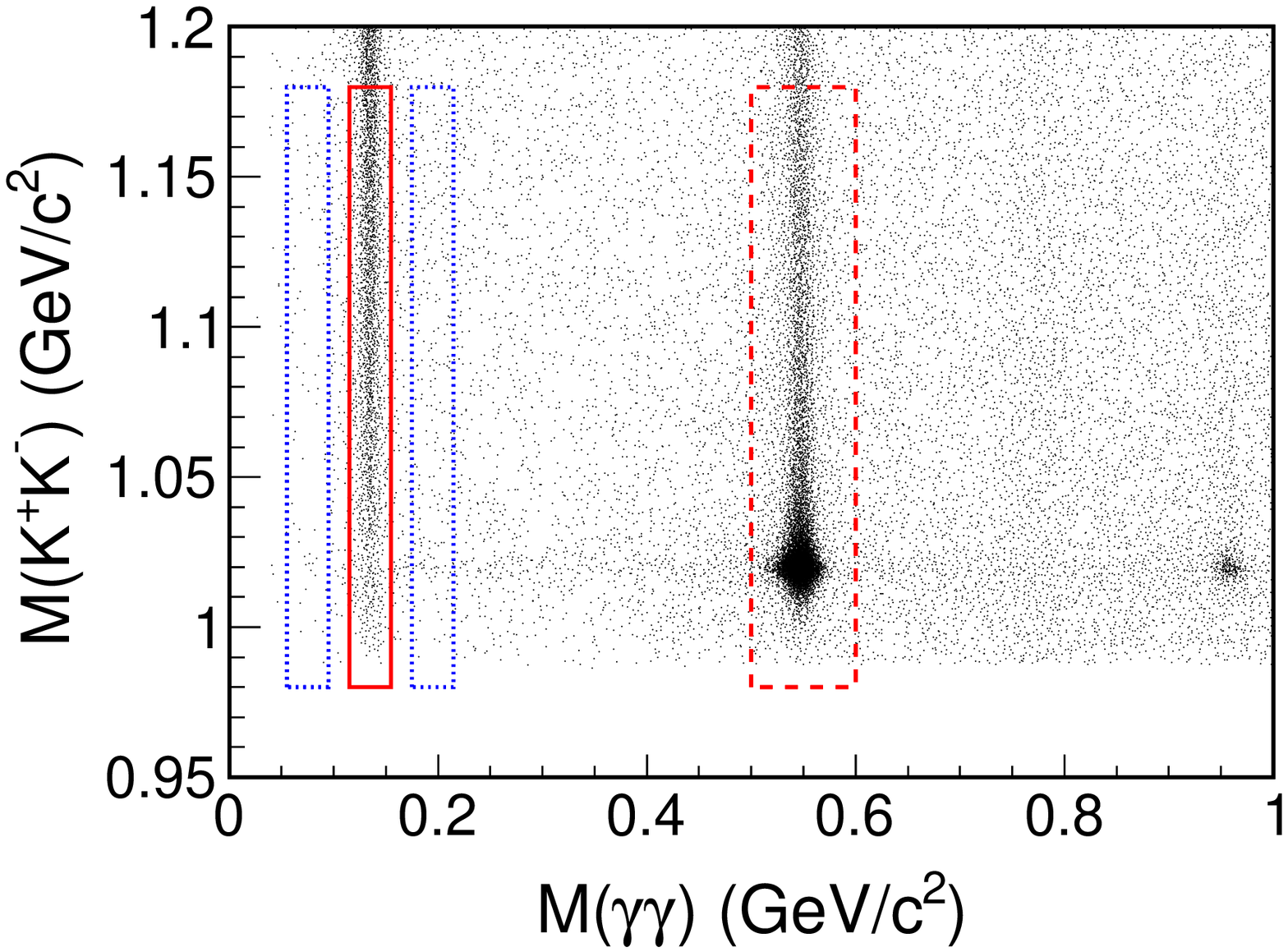}
\put(-150, 105){\textbf{(a)}}\\
\includegraphics[width=0.4\textwidth]{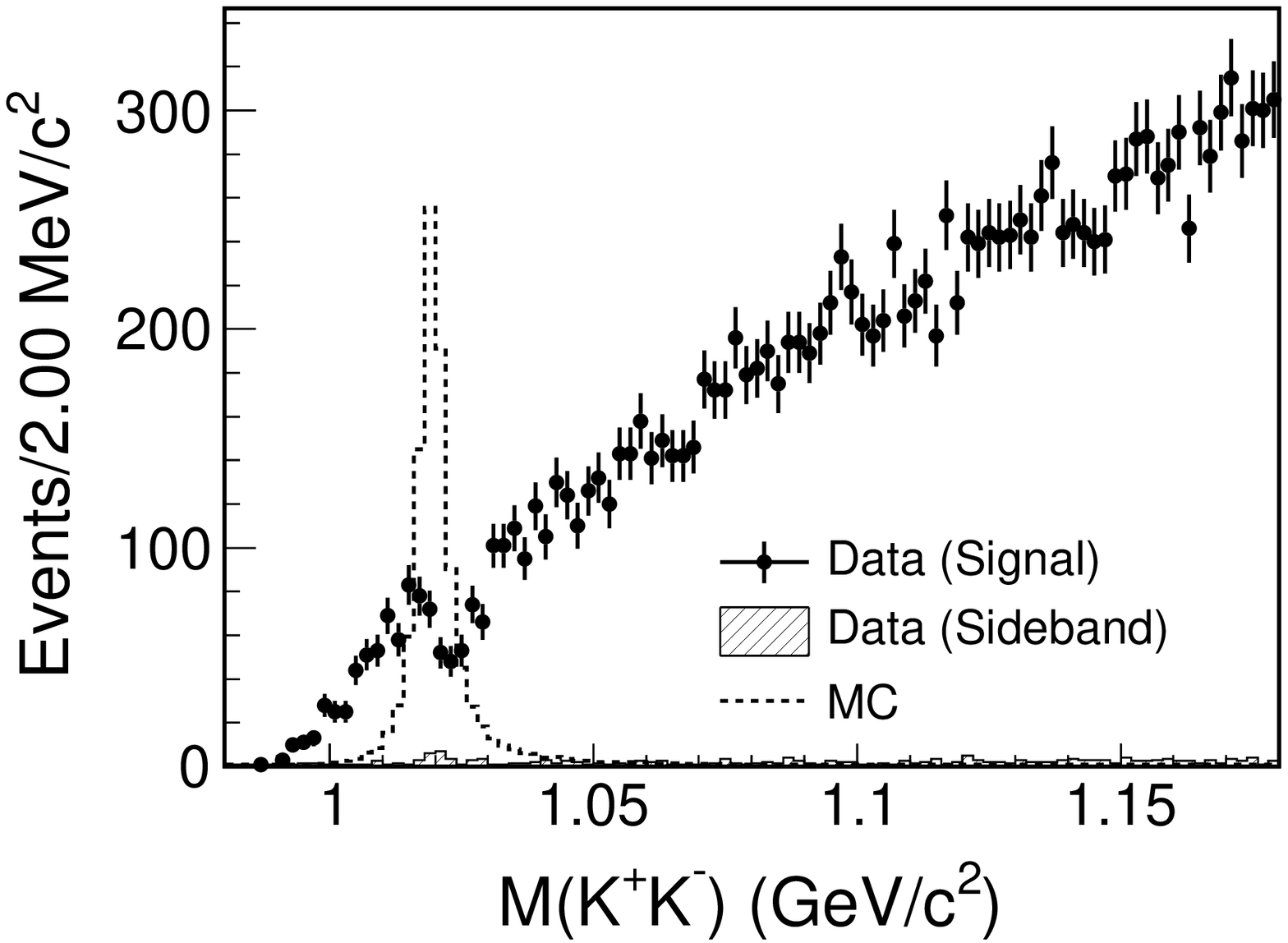}
\put(-150, 105){\textbf{(b)}}\\
  \caption{(a) Scatter plot of $M(\Kp\Km)$ versus
  $M(\gamma\gamma)$. The red solid and blue dotted boxes are the $\piz$
  signal and sideband regions. The red dashed  box indicates
  $\jpsi \to \phi \eta$. (b) $\Kp\Km$ mass spectrum, where the dots
  with error bars are events in the $\piz$ signal region; the hatched
  histogram are events from the $\piz$ sidebands; and the dashed
  histogram is MC simulation of $\jpsi \to \phi \piz$ with arbitrary
  normalization.}\label{fig:Mphi_Mpiz}
\end{figure}

After the above selection, the scatter plot of $M(\Kp\Km)$ versus
$M(\gamma\gamma)$ (Fig.~\ref{fig:Mphi_Mpiz} (a)) shows two clear
clusters corresponding to $\phi\eta$ and $\phi\eta^\prime$ and two
bands corresponding to $\Kp\Km\piz$ and $\Kp\Km\eta$ , but no evident
accumulation of events for $\phi\piz$. To investigate the $M(\Kp\Km)$
spectrum of $\Kp\Km\piz$ events, we select events where the
$\gamma\gamma$ invariant mass is in the $\piz$ mass region
$0.115<M(\gamma\gamma)<0.155$~GeV/$c^2$.  The $M(\Kp\Km)$ distribution
for these events is shown in Fig.~\ref{fig:Mphi_Mpiz} (b),
where a clear structure around the $\phi$ mass is seen.

\begin{table}
\center{}
\caption{\label{tab:bkg} Background analysis for the decay $\jpsi \to \phi\piz$.}
\begin{ruledtabular}
\begin{tabular}{l c}
Type & Reactions\\
\hline
Coherent & $\jpsi \to \Kp\Km\piz$ \\
$\phi$ peaking  & $e^+e^-\to\gamma_{ISR}\phi$, $\jpsi \to \phi \piz\piz/\phi\gamma\gamma$ \\
$\piz$ peaking & $\jpsi \to \gamma \eta_c(1S) \to \gamma \Kp\Km\piz$\\
Other & $\jpsi \to \gamma\Kp\Km/\gamma\piz\Kp\Km/\piz\piz\Kp\Km$
\end{tabular}
\end{ruledtabular}
\end{table}

Studies were performed using both MC events and data to investigate whether
the structure around 1.02~GeV/$c^2$ could be background related.
We analyze a MC sample of $1.2 \times 10^{9}$ $J/\psi$ inclusive
decays, in which the known decay modes were generated by BesEvtGen~\cite{EvtGen,BesMC}
with measured branching fractions~\cite{PDG}
while unknown decays were generated by Lund-Charm~\cite{Lund}.
The dominant background events are found to be from
$J/\psi\rightarrow K^+K^-\pi^0$ with the intermediate states decaying
into $K^{\pm}\piz$ and $\Kp\Km$,
which is coherent for the decay $\jpsi \to \phi\piz$.
A partial wave analysis, not including $\jpsi \to \phi\piz$ but considering the
interference of all intermediate states, yields a smooth distribution
with $\Kp\Km$ mass below 1.2~GeV/$c^2$.
The incoherent background can be categorized into three classes as follows. (1) The $\phi$ peaking background: $e^+e^-\to \gamma_{ISR}\phi$ and $\jpsi \to \phi\piz\piz/\phi\gamma\gamma$. The former background is studied using data taken at energies far from any charmonium resonance and the latter ones are studied by exclusive MC samples. The studies show that these background events can be compensated by the $\piz$ mass sideband events, which
are defined as $0.055<M(\gamma\gamma)<0.095$~GeV/$c^2$ and
$0.175<M(\gamma\gamma)<0.215$~GeV/$c^2$.
(2) The $\piz$ peaking background: $\jpsi \to \gamma \eta_c(1S) \to \gamma\Kp\Km\piz$. This background cannot be taken into account by the $\piz$ mass sideband events. From MC simulations, the ratio between the number of this background events and the number of the coherent background events in the $\piz$ mass region is 0.5\%. As little is known about the possible intermediate states, we neglect this background and consider the related systematic uncertainty. (3) The non-$\phi$ and non-$\piz$ background are dominated by the decays $\jpsi \to \gamma\Kp\Km$, $\gamma\piz\Kp\Km$ and $\piz\piz\Kp\Km$ with various intermediate states. MC simulations show they can be subtracted by the $\piz$ mass sideband events. All background types are summarized in Table~\ref{tab:bkg}.
Through the studies above, none of these background events produce a structure in the $K^+K^-$ mass spectrum. In addition, the detection
efficiency as a function of $M(K^+K^-)$, obtained from the MC
simulation and taking into account the angular
distributions~\cite{angular}, is also smooth over the $K^+K^-$ mass
region, with no structure in the region of the $\phi$ signal.

A possible explanation for the structure in the $M(\Kp\Km)$ spectrum is
interference between $\jpsi \to \phi \piz$ and other processes with
the same final state. We have verified this using a statistical hypothesis
test~\cite{James,Cowan}. In the null hypothesis without
$J/\psi\to\phi\piz$, a second-order polynomial function,
defined as $F_{H_0} = P(m) \equiv c_0 + c_1 m + c_2 m^2$, is used in
the fit to describe the data after subtraction of the $\piz$-sideband
events. The positive hypothesis is characterized by a
two-component function ($F_{H_1}$), in which the model is a coherent sum of
a relativistic Breit-Wigner resonance and the second-order polynomial function, convoluted with a Gaussian
function $G(m, \sigma_m)$ to take into account the mass resolution,
$\sigma_m$.

\begin{equation}{\label{eq:dNdm}}
F_{H_1} = |\sqrt{P(m)/\Phi(m)} + A_{\phi}(m)|^2 \Phi(m)\otimes G(m, \sigma_m) \: ,
\end{equation}
where
\begin{equation}{\label{eq:Aphi}}
A_\phi(m) = \frac{\sqrt{R}e^{i\delta}p_\phi(m)p_K(m)
}{m^2-m_0^2 + i m\Gamma(m)}\frac{B(p_\phi(m))}{B(p_\phi(m_0))}\frac{B(p_K(m))}{B(p_K(m_0))},
\end{equation}
with
\begin{equation}{\label{eq:GammaB}}
\Gamma(m) \equiv \left( \frac{p_K(m)}{p_K(m_0)}\right)^3 \frac{m_0}{m} \frac{B(p_K(m))}{B(p_K(m_0))}\Gamma_0 \: .
\end{equation}
Here, $m$ is the $\Kp\Km$ invariant mass. $m_0$ and $\Gamma_0$ are the nominal mass and decay width of the
$\phi$~\cite{PDG}. $p_\phi(m)$ ($p_K(m)$) is the momentum of the $\phi$ ($K$) in the frame of $\jpsi$
($\phi$) with the mass of $\phi$ being $m$. $\Phi(m) = p_\phi(m)p_K(m)$ is the phase space factor.
$B(p)$, defined as $B(p)\equiv 1/\sqrt{1 + (rp)^2}$, is the Blatt-Weisskopf penetration
form factor~\cite{Blatt} with the meson radius $r$ being $3$
GeV$^{-1}$. $R$ and $\delta$ represents the magnitude and relative phase angle respectively
for the contribution of the $\phi$ resonance. Omitting the convolution, $F_{H_1}$ can be expanded to be
\begin{equation}\label{eq:FH1}
P(m) + |A_{\phi}(m)|^2\Phi(m) + 2\sqrt{P(m)\Phi(m)}\Re A_{\phi}(m) \: ,
\end{equation}
where the first term is the non-$\phi$ contribution from the decay $\jpsi \to \Kp\Km\piz$; the second term is the $\phi$ resonance from the decay $\jpsi \to \phi\piz$; and the third term is their interference. Here, $\Re A_{\phi}(m)$ denotes the real part of $A_{\phi}(m)$.

MC simulations show that the $\Kp\Km$ mass resolutions are essentially
the same for $\jpsi$ decaying to $\phi\piz$ and $\phi\eta$ with
$\piz/\eta\to\gamma\gamma$. We obtain $\sigma_m = (1.00\pm0.02)$
MeV/$c^2$ by performing an unbinned likelihood fit to the $M(\Kp\Km)$
spectrum of $\jpsi \to \phi\eta$ with $0.50<M(\gamma\gamma)<0.60$
GeV/$c^2$, shown in the red dashed box in
Fig.~\ref{fig:Mphi_Mpiz} (a). The same Breit-Wigner formula convoluted
with a Gaussian function is used to describe the $\phi$ signal, while a
second-order polynomial is used to describe the background.

\begin{figure}
\includegraphics[width=0.4\textwidth]{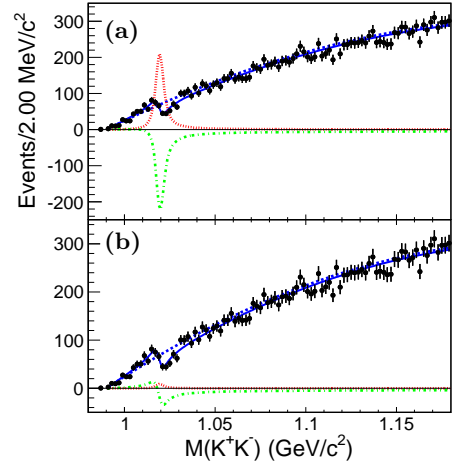}
\put(-150, 160){\textbf{(a)}}
\put(-150, 80){\textbf{(b)}}
  \caption{Fit to $M(\Kp\Km)$ spectrum after sideband subtraction for
  Solution I (a) and Solution II (b). The red dotted curve denotes the
  $\phi$ resonance; the blue dashed curve is the non-$\phi$
  contribution; the green dot-dashed curve represents their interference;
  and the blue solid curve is the sum of them.}\label{fig:Mphi_solution}
\end{figure}

After subtracting the incoherent background events estimated with
$\piz$ sidebands, a maximum likelihood fit is performed to the
$M(K^+K^-)$ distribution under the positive hypothesis.
Two solutions with two different phase angles between the $\phi$ resonance and the non-$\phi$ contributions are found.
The final fits, including the individual contributions of each components, are illustrated in
Fig.~\ref{fig:Mphi_solution}~(a) and Fig.~\ref{fig:Mphi_solution}~(b),
while the signal yields and the relative phase angles are summarized in Table~\ref{tab:fit}.
In Fig.~\ref{fig:Mphi_solution}~(a) and Fig.~\ref{fig:Mphi_solution}~(b),
the blue dashed curve is the non-$\phi$ contribution (the first term in Eq.~\ref{eq:FH1});
the red dotted curve denotes the $\phi$ resonance (the second term in Eq.~\ref{eq:FH1});
the green dot-dashed curve represents their interference (the third term in Eq.~\ref{eq:FH1});
and the blue solid curve is the sum of them.
The signal yield $N^\text{sig}$ in Table~\ref{tab:fit} is calculated by integrating the function of the $\phi$ resonance over the fit range.
The statistical significance is determined by
the change of the log likelihood value and the number of degrees of
freedom in the fit with and without the $\phi$ signal~\cite{James,Wilks}.
Both solutions have a statistical significance of 6.4$\sigma$,
which means that they provide identically good descriptions of data.

\begin{table}
\center{}
\caption{\label{tab:fit} Fit results. $N^\text{sig}$ is the fitted number
of signal events (from the parameter $R$). $\delta$ is the relative
phase. $2\Delta \log \mathcal{L}$ is $2$ times the difference of the
log-likelihood value with and without $\phi$ signal, while $N_f$ is
the change of the number of degrees of freedom. $Z$ is the statistical
significance.}
\begin{ruledtabular}
\begin{tabular}{c c c c c}
 Solution &$N^\text{sig}$ & $\delta$ & $2\Delta\log \mathcal{L} /N_f$ & $Z$\\
\hline
 I  & $\NobsA$ & $\phaseA$ & $45.8/2$ &  $6.4\sigma$\\
 II  & $\NobsB$ & $\phaseB$ & $45.8/2$ &  $6.4\sigma$\\
\end{tabular}
\end{ruledtabular}
\end{table}

The non-resonant quantum electrodynamics (QED) contribution is
estimated in two ways. One way is by analyzing data taken at energies
far from any resonance, namely at $3.05$, $3.06$, $3.08$,
$3.083$, $3.090$ and $3.65$~GeV. The other way is to use data
from the $\psi(3770)$ resonance, assuming that the possible contribution $\psi(3770)\to\phi\piz$ is negligible.
The selection criteria are the same
except for the required 4-momenta in the kinematic
fit.
Neither sample shows significant $\phi\piz$ events. With a
simultaneous fit, we obtain the QED contribution to the signal yield
$N_{\phi\piz}^\text{con}(3.097)<5.8$ at the $90\%$ C.L., normalized
according to the luminosity and efficiency and assuming the cross
section is proportional to $1/s$ with $s$ being the square of the
c.m. energy. Thus we neglect the non-resonant QED contribution and use
the upper limit of $N_{\phi\piz}^\text{con}(3.097)$ to estimate a
systematic uncertainty from this assumption.

With the detection efficiency, $(45.1\pm0.2)\%$, obtained from the MC
simulation, the branching fractions of $\jpsi \to \phi \piz$ are
calculated to be $\solutionAs$ for Solution I and $\solutionBs$ for
Solution II, where the errors are statistical only.

The sources of systematic uncertainty and their corresponding contributions
to the measurement of the branching fraction are summarized in
Table~\ref{tab:se_summary}. The tracking efficiency of charged kaons
is studied using a high-purity control sample $\jpsi \to
K_S^{0}K^\pm\pi^\mp$, while the photon detection efficiency is
investigated based on a clean sample of $\jpsi \to \rho\pi$.  The
differences between data and MC simulation are $1\%$ for each charged
track and $1\%$ for each photon. The $\piz$ selection efficiency is
studied with the sample $\jpsi \to \rho\pi$, and MC simulation agrees with data within $0.6\%$.
The particle identification efficiency is
studied with the sample $\jpsi \to \phi\eta \to
\Kp\Km\gamma\gamma$. The efficiency difference between data and MC is
0.5\%.  To estimate the uncertainty associated with the kinematic
constraint, a control sample of $\jpsi \to
\phi\eta\to\Kp\Km\gamma\gamma$ is selected without a kinematic fit. The
efficiency is the ratio of the signal yields with and without the
kinematic requirement $\chi^2(4C)<30$. The difference between data and
MC, $3.2\%$, is assigned as the systematic uncertainty.  For the
uncertainties from the fit, alternative fits are performed by
varying the bin size and fit ranges. In addition, we also consider
the effect from the parameterization of the function ($F_{H_0}$) for
the null hypothesis, the relative phase angle $\delta$ and the decay width $\Gamma(m)$. We repeat fits
parameterizing $F_{H_0}$ with a third-order polynomial and extending
$\delta$ in Eq.~\ref{eq:dNdm} to be $\delta + \delta_1
\frac{m-m_0}{\Gamma_0} + \delta_2 (\frac{m-m_0}{\Gamma_0})^2$ with two
more parameters $\delta_1$ and $\delta_2$. Assuming the modes $\phi\to K^+K^-/K_SK_L$ have the same branching fraction 50\%, we also perform a fit replacing $\Gamma(m)$ with $\Gamma_{K^+}(m)\times 50\% + \Gamma_{K^{0}}(m)\times 50\%$,
where $\Gamma_{K^+/K^0}(m)$ is Eq.~\ref{eq:GammaB} using the mass of $K^+/K^0$.
The yield difference with
respect to the nominal fit is taken as the systematic uncertainty due
to the parameterization.  The
mass resolution, $\sigma_m = 1.00 \pm 0.02$~MeV/$c^2$, is determined
from $J/\psi\rightarrow\phi\eta$. Varying $\sigma_m$ within
$\pm 0.02$~MeV/$c^2$ in the fit, the signal yield difference compared
to the nominal fit is less than $1\%$. The QED contribution is
neglected and the uncertainty for $N_{\phi\piz}^\text{con}(3.097)$ is taken
as $5.8$ as stated above. It contributes a systematic uncertainty of $0.7\%$
($16.4\%$) for Solution I (II), ignoring the possible interference
between the QED process and $\jpsi$ resonance decay.  The mass and
width of the $\phi$ meson have been fixed to their world
averages~\cite{PDG}. Changing them with $1\sigma$ uncertainty, the
signal yield difference is taken as the systematic uncertainty. The meson
radius $r$ is 3~GeV$^{-1}$ in the nominal fit. We change it from 1
GeV$^{-1}$ to 5~GeV$^{-1}$, and the largest signal yield difference is
2.3\% (3.0\%) for Solution I (II).
In the fit, the $\piz$ peaking background $\jpsi \to \gamma \eta_c(1S) \to \gamma\Kp\Km\piz$ is neglected. These background events can be subtracted by a MC simulation normalized according to the relevant branching fractions~\cite{PDG} and the efficiency. The signal yield difference is 1.1\% (3.8\%) for Solution I (II).
We also consider the
uncertainties from the number of $\jpsi$ events and the branching
fraction of $\phi \to \Kp\Km$. The total systematic uncertainty in
Table~\ref{tab:se_summary} is the quadratic sum of the individual
ones, assuming they are independent.

\begin{table}
\center{}
\caption{\label{tab:se_summary} Summary of branching fraction
systematic uncertainties (in $\%$).}
\begin{tabular}{l  c  c}
\hline\hline
 Source & Solution I & Solution II \\
 \hline
MDC tracking & 2.0 & 2.0\\
Photon detection & 2.0 & 2.0\\
Particle identification & 0.5 & 0.5 \\
$\piz$'s selection & 0.6 & 0.6\\
Kinematic fit & 3.2 & 3.2\\
Bin size & 1.0 & 6.5\\
Fit range & 1.0 & 15.3\\
Mass resolution & 0.1 & 0.4\\
Parameterization of $F_{H_0}$ & 0 & 1.9\\
Parameterization of $\delta$ & 0.9 & 1.6\\
Parameterization of $\Gamma(m)$ & 0.1 & 0.0 \\
QED continuum & 0.7 & 16.4\\
The mass and width of $\phi$ & 0.8 & 0.1 \\
The meson radius $r$ & 2.3 & 3.0 \\
$\piz$ peaking background & 1.1 & 3.8 \\
Number of $\jpsi$ & 0.8 & 0.8\\
Uncertainty of $\BR(\phi\to \Kp\Km)$ & 1.0 & 1.0\\
\hline
Total & 5.5 & 24.4\\
\hline\hline
\end{tabular}
\end{table}

\begin{figure}
\includegraphics[width=0.4\textwidth]{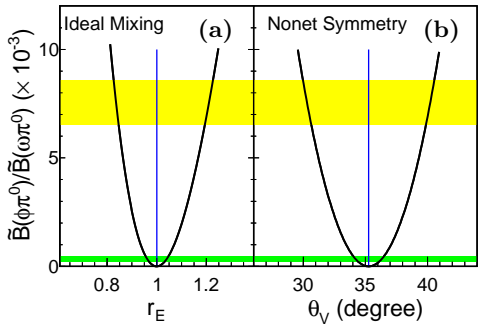}
\put(-115, 110){\textbf{(a)}}
\put(-30, 110){\textbf{(b)}}
\caption{Dependence of the reduced branching fraction ratio
$\tilde{\BR}(\phi\piz)/\tilde{\BR}(\omega \piz)$ (a) on the nonet
symmetry breaking strength $r_E$ assuming $\omega-\phi$ ideal mixing
and (b) on the mixing angle $\theta_V$ assuming nonet symmetry. The
yellow (green) box represents Solution I (II). The blue line
represents the nonet symmetry value in (a) and the ideal mixing angle
in (b). }\label{fig:theory}
\end{figure}

In summary, based on 1.31 billion $\jpsi$ events collected with the
BESIII detector, we perform an analysis of the decay $\jpsi \to \phi
\piz \to \Kp \Km \gamma\gamma$ and find a structure around $1.02$~GeV/$c^2$
in the $\Kp\Km$ invariant mass spectrum. It can be interpreted as
interference of $\jpsi \to \phi \piz$ with other processes decaying to
the same final state. The fit yields two possible solutions and thus
two branching fractions, $\solutionA$ and
$\solutionB$.

Ref.~\cite{Haber} provides a model-independent relation,
$\tilde{\BR}(\phi\piz)/\tilde{\BR}(\omega \piz) =
(r_E\tan\theta_V-1/\sqrt{2})^2/(\tan\theta_V/\sqrt{2}+r_E)^2$. Here
$\tilde{\BR}(VP)\equiv \BR(VP)/p_V^3$ is the reduced branching
fraction of the decay $\jpsi \to VP$, and $p_V$ is the momentum of the
vector meson in the rest frame of $\jpsi$; $\theta_V$ is the
$\omega-\phi$ mixing angle; $r_E$ is a dimensionless parameter
accounting for nonet symmetry breaking in the electromagnetic sector
and $r_E=1$ corresponds to nonet symmetry. We have used $\BR(\jpsi \to
\omega \piz) = (4.5\pm0.5)\times10^{-4}$~\cite{PDG}. If $\omega-\phi$
are mixed ideally, namely
$\theta_V=\theta_V^\text{ideal}\equiv\arctan\frac{1}{\sqrt{2}}$, the nonet
symmetry breaking strength is $\delta_E\equiv r_E-1 = (+21.0\pm1.6)\%$
or $(-16.4 \pm 1.0)\%$ ($(+3.9 \pm 0.8)\%$ or $(-3.7 \pm 0.7)\%$) for
Solution I (II), illustrated in Fig.~\ref{fig:theory} (a). On the
other hand, we obtain $\phi_V \equiv
|\theta_V-\theta_V^\text{ideal}|=4.97^\circ\pm0.33^\circ$
($1.03^\circ\pm0.19^\circ$) for Solution I (II) assuming nonet
symmetry, shown in Fig.~\ref{fig:theory} (b). However, $\phi_V$ is
found to be $3.84^\circ$ from the quadratic mass formulae~\cite{PDG}
and $3.34^\circ \pm 0.09^\circ$ from a global fit to the radiative
transitions of light mesons~\cite{KLOE}. The $\phi_V$ values do not
agree with either solution. This is the first indication that nonet
symmetry~\cite{Haber} is broken and the doubly OZI-suppression process contributes
in $\jpsi$ electromagnetic decays.

The BESIII collaboration thanks the staff of BEPCII and the IHEP computing center for their strong support. This work is supported in part by National Key Basic Research Program of China under Contract No. 2015CB856700; National Natural Science Foundation of China (NSFC) under Contracts Nos. 11125525, 11235011, 11322544, 11335008, 11425524; the Chinese Academy of Sciences (CAS) Large-Scale Scientific Facility Program; Joint Large-Scale Scientific Facility Funds of the NSFC and CAS under Contracts Nos. 11179007, U1232201, U1332201; CAS under Contracts Nos. KJCX2-YW-N29, KJCX2-YW-N45; 100 Talents Program of CAS; INPAC and Shanghai Key Laboratory for Particle Physics and Cosmology; German Research Foundation DFG under Contract No. Collaborative Research Center CRC-1044; Istituto Nazionale di Fisica Nucleare, Italy; Ministry of Development of Turkey under Contract No. DPT2006K-120470; Russian Foundation for Basic Research under Contract No. 14-07-91152; U. S. Department of Energy under Contracts Nos. DE-FG02-04ER41291, DE-FG02-05ER41374, DE-FG02-94ER40823, DESC0010118; U.S. National Science Foundation; University of Groningen (RuG) and the Helmholtzzentrum fuer Schwerionenforschung GmbH (GSI), Darmstadt; WCU Program of National Research Foundation of Korea under Contract No. R32-2008-000-10155-0


\begin{thebibliography}{99}
\bibitem{angular}
$\frac{d^3\sigma}{d\cos\theta_\phi d\cos\theta_1 d\phi_1} =
\sin^2 \theta_1 [1 + \cos^2\theta_\phi + \sin^2\theta_\phi \cos (2\phi_1)]$, where $\theta_\phi$ is the angle between the $\phi$ meson and the positron direction. $\theta_1$ and $\phi_1$ are the polar and azimuthal angles of the momentum of $K$ with respect to the helicity direction of the $\phi$ meson.
\end{thebibliography}
\end{document}